\documentclass[12pt]{article}

\newcommand{\epb}{\mbox{b}}
\newcommand{\epba}{\overline{\mbox{b}}}
\newcommand{\epc}{\mbox{c}}
\newcommand{\epca}{\overline{\mbox{c}}}
\newcommand{\epd}{\mbox{d}}
\newcommand{\epda}{\overline{\mbox{d}}}
\newcommand{\epe}{\mbox{e}^-}
\newcommand{\epea}{\mbox{e}^+}

\newcommand{\epq}{\mbox{q}}

\newcommand{\epqa}{\overline{\mbox{q}}}

\newcommand{\epu}{\mbox{u}}
\newcommand{\epua}{\overline{\mbox{u}}}
\newcommand{\ferm}{\mbox{f}}

\newcommand{\ggn}{\mbox{g}}

\newcommand{\hp}{\mbox{p}}
\newcommand{\hpa}{\overline{\mbox{p}}}
\newcommand{\mba}{\overline{\mbox{B}}}
\newcommand{\mbn}{\mbox{B}^0}
\newcommand{\mbna}{\overline{\mbox{B}} \raisebox{1ex}{\scriptsize 0}}
\newcommand{\mbp}{\mbox{B}^+}
\newcommand{\mpim}{\pi^-}
\newcommand{\mpip}{\pi^+}

\newcommand{\bibl}[5]
	{#1, {\it #2} {\bf #3} #5 (#4)}
\newcommand{\cebe}{\begin{center}}
\newcommand{\ceen}{\end{center}}
\newcommand{\debe}{\begin{description} \vspace{-2ex}}
\newcommand{\deen}{\end{description}}
\newcommand{\eabe}{\begin{eqnarray}}
\newcommand{\eaen}{\end{eqnarray}}
\newcommand{\epem}{$\mbox{e}^+\mbox{e}^-$}
\newcommand{\eqbe}{\begin{equation}}
\newcommand{\eqen}{\end{equation}}
\newcommand{\etal}{{\em et al.} }

\newcommand{\ra}{\rightarrow}

\input{psfig}

\begin{document}

\begin{titlepage}
\begin{flushright}
  LU TP 96-11 \\
  April 1996
\end{flushright}
\vspace{25mm}
\begin{center}
  \Large
  {\bf $\epb \epba$-fragmentation and B$\mathbf{\pi}$ correlations} \\ 
  \normalsize
  \vspace{12mm}
  Jari H\"akkinen\footnote{jari@thep.lu.se} \\
  Department of Theoretical Physics, Lund University, \\
  S\"olvegatan 14A, SE-223 62 Lund, Sweden
\end{center}
\vspace{20mm}

\noindent {\bf Abstract:} \\
We show that experimental data are in very good agreement with predictions from the string fragmentation model by Bowler and Morris. We present a physical interpretation and discuss the relation to results obtained from Perturbative QCD and Local Hadron Parton Duality (LPHD). We also present implications for B$\pi$ correlations and the possibility to use these as a tag to study CP violation in B decays.
\end{titlepage}

\section{Introduction} \vspace{-2ex}
An understanding of hadronization in b-quark jets is important for several reasons.\vspace{-3ex} 
\begin{itemize}
  \item It can help to understand the confinement mechanism. \vspace{-2ex}
  \item A good description of jet fragmentation is essential to reconstruct an 
	underlying hard parton interaction. Thus e.g. a measurement of the 
	reaction $\epea \epe \ra \mbox{W}^+ \mbox{W}^- \ra \epq_1 \epqa_2 \epq_3
	\epqa_4$ at LEP2 is important for determination of the W mass, which 
	can give information about virtual corrections involving the Higgs 
	particle. \vspace{-2ex}
  \item As proposed in~\cite{mg92} particle correlations in b and $\epba$ jets 
	can be used as a tag which e.g. can improve a determination of CP
	violation in B decay.
\end{itemize}

The Lund string fragmentation model~\cite{ba83} (implemented in the Jetset MC~\cite{ts94}) and the cluster fragmentation model~\cite{bw84} (implemented in the Herwig MC~\cite{gm92}) have been frequently used by experimental groups to describe the hadronization process in the analysis of their data. Previous official versions (up to v7.3) of the Jetset MC have in the default version given a too hard B meson spectrum, and in consequence too low total multiplicity in $\epb \epba$ events. To describe the data, experimentalists have instead often used the Peterson et al. model~\cite{cp83} for the B meson momentum, combined with string fragmentation for the remainder of the jet. That model contains one free parameter which can be adjusted to the experimental data. The two most essential experimental observables are the average B-meson energy (or $\overline{x}_B$) and the average total charged multiplicity $\overline{n}$. In the Jetset MC it is assumed that after separation of the B hadron, the rest of the jet corresponds to a light quark jet with the remaining energy, and therefore there is a correlation between $\overline{x}_B$ and $\overline{n}$. It is noteworthy that this correlation agrees well with data. Bowler and Morris~\cite{mb81} have proposed a modification of the initial Lund Model for the fragmentation of heavy b or c quarks. This model has no free parameter, and therefore gives a definite prediction. In the present paper we will demonstrate that the model\vspace{-2ex}
\begin{itemize}
\item agrees well with expectations from a string dynamics scenario. \vspace{-2ex}
\item agrees qualitatively with expectations from perturbative QCD and local parton hadron duality or cluster fragmentation. \vspace{-2ex}
\item agrees very well with experimental data.
\end{itemize}

Equipped with a description of b fragmentation, we will in this paper also study the B$\pi$ correlations and estimate the efficiency of the tagging method proposed in ref~\cite{mg92} for a study of CP violation in B decays.

\section{b jet fragmentation} \vspace{-2ex}
The Lund string fragmentation model is implemented in Jetset to take care of the fragmentation of $\epq \epqa$ strings. The main idea in the fragmentation of the original $\epq \epqa$ pair, ie. $\epea \epe \ra Z^0, \gamma \ra \epq \epqa$, is that new quark--anti-quark (diquark--anti-diquark) pairs are produced in the colour field stretched between the original quark pair. The constituents of a new pair obtain their transverse mass through the tunneling process needed to create them~\cite{ba83}. There is a flavour ordering present throughout the string so that the flavour quantum number is preserved. Details of how these properties are given to hadrons produced in Jetset will not concern us here. Instead we are going to study the energy and longitudinal momentum distribution of the produced hadrons.

\begin{figure}[t,b]
  \hbox{\vbox{
    \begin{center}
    \mbox{\psfig{figure=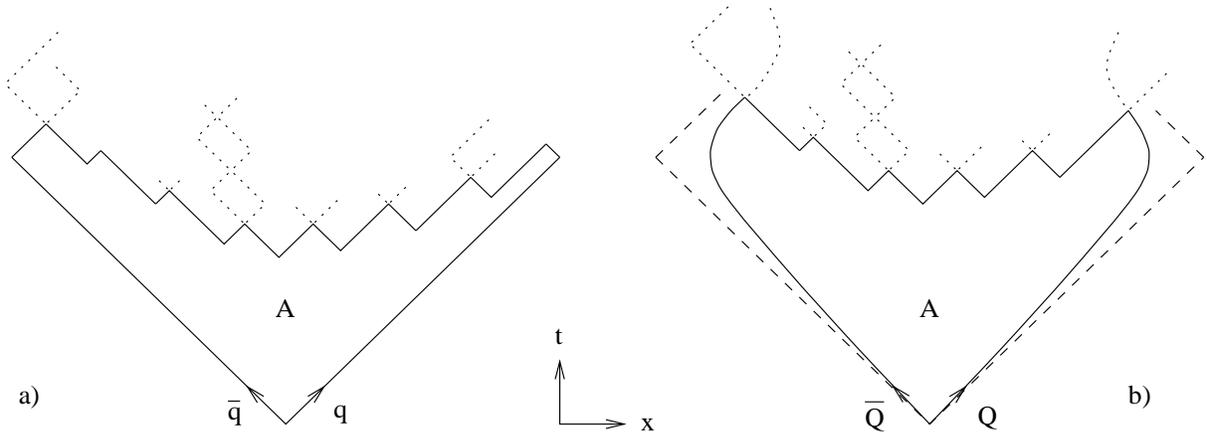,width=16cm}}
    \end{center}
  }}
  \caption{\em Quark trajectories in the Lund model. Light quarks move along light cones \em a), \em while heavy quarks move along hyperbolae \em b). \em The finite mass of heavy quarks decreases the area bounded by the solid lines resulting in softer B spectrum.}
  \label{f:phasespace}
\end{figure}

A basic assumption in the Lund model is ``left-right symmetry'', which means that the fragmentation process should look the same irrespective to which end of the string we start from in an iterative process. From this assumption follows that the probability $dP$ to obtain a definite final state with hadron momenta $p_i$ (being 1+1 dimensional vectors if we neglect the transverse dimensions) with masses $m_i$ is given by~\cite{ba83:2}
\eqbe
dP \propto \prod_i N_i d^2p_i \delta(p^2_i - m^2_i) \delta(\sum_k p_k -p_{total}) \exp(-bA)
\label{e:area}
\eqen
where $A$ is the area indicated in Fig~\ref{f:phasespace}.
It is suggestive that this expression is the product of a phase space factor and the exponent of an area, which can be interpreted as (the imaginary part of) an action and is similar to a Wilson loop integral. 

The distribution in Eq~(\ref{e:area}) can be generated iteratively starting from one end of the quark--anti-quark system. With the other properties fixed, the probability to give a hadron a fraction $z$ of the available energy is given from a splitting function,
\eqbe
f(z) \propto \frac{1}{z}z^{a_i} \left( \frac{1-z}{z} \right)^{a_k}
\exp \left(- \frac{b m^2_{\perp}}{z} \right).
\label{e:original}
\eqen
The parameters $a_i$ could in principle depend on the flavour of the associated quark--anti-quark pair. In Eq~(\ref{e:original}) the index $i$ corresponds to the flavour of the previously produced $\epq \epqa$ pair and $k$ to the latest produced flavour pair. Since experimental data seems to need the use of only two different $a$-values in connection with quark--anti-quark and diquark--anti-diquark production respectively (the latter leading to baryon--anti-baryon production), two different $a$-values have been used in most applications.

We also note that in the string model it is assumed that heavy quarks like c or b can only be produced perturbatively, either directly from the initial $\gamma$ or Z, or in the process $\ggn \ra \epq \epqa$, and not in the soft string breaking process. Thus a b-quark is always at the end of a string, and the argument based on left-right symmetry is really not applicable for the B hadrons. The fragmentation will always be left-right symmetric whatever energy is given to the B and the $\mba$ at the ends of the string. A modification of the splitting function in Eq~(\ref{e:original}), based on the physical ideas behind the string model, was proposed by Bowler and Morris in~\cite{mb81}. If the initial $\epq \epqa$ pair is a heavy $\epc \epca$ or $\epb \epba$ pair, these quarks do not move along the light cones but instead along hyperbolae in the $x-t$ diagram as shown in Fig~\ref{f:phasespace}b. It is then natural to insert in Eq~(\ref{e:area}) the area bounded by the quark trajectory as indicated in Fig~\ref{f:phasespace}b~\cite{mb81}. The result is a softer spectrum for the leading B meson, which is well approximated by the expression
\eqbe
f(z) \propto \frac{1}{z^{1+r_Q b M^2_Q}} z^{a_i} \left( \frac{1-z}{z} \right)^{a_k}
\exp \left(- \frac{b m^2_{\perp}}{z} \right).
\label{e:modified}
\eqen
Here $r_Q$ is predicted to be equal to 1, but is introduced as a parameter in Jetset. This is done to make it possible to control the effect of the modification.

\begin{figure}[t,b]
  \hbox{\vbox{
    \begin{center}
       \mbox{
	  \psfig{figure=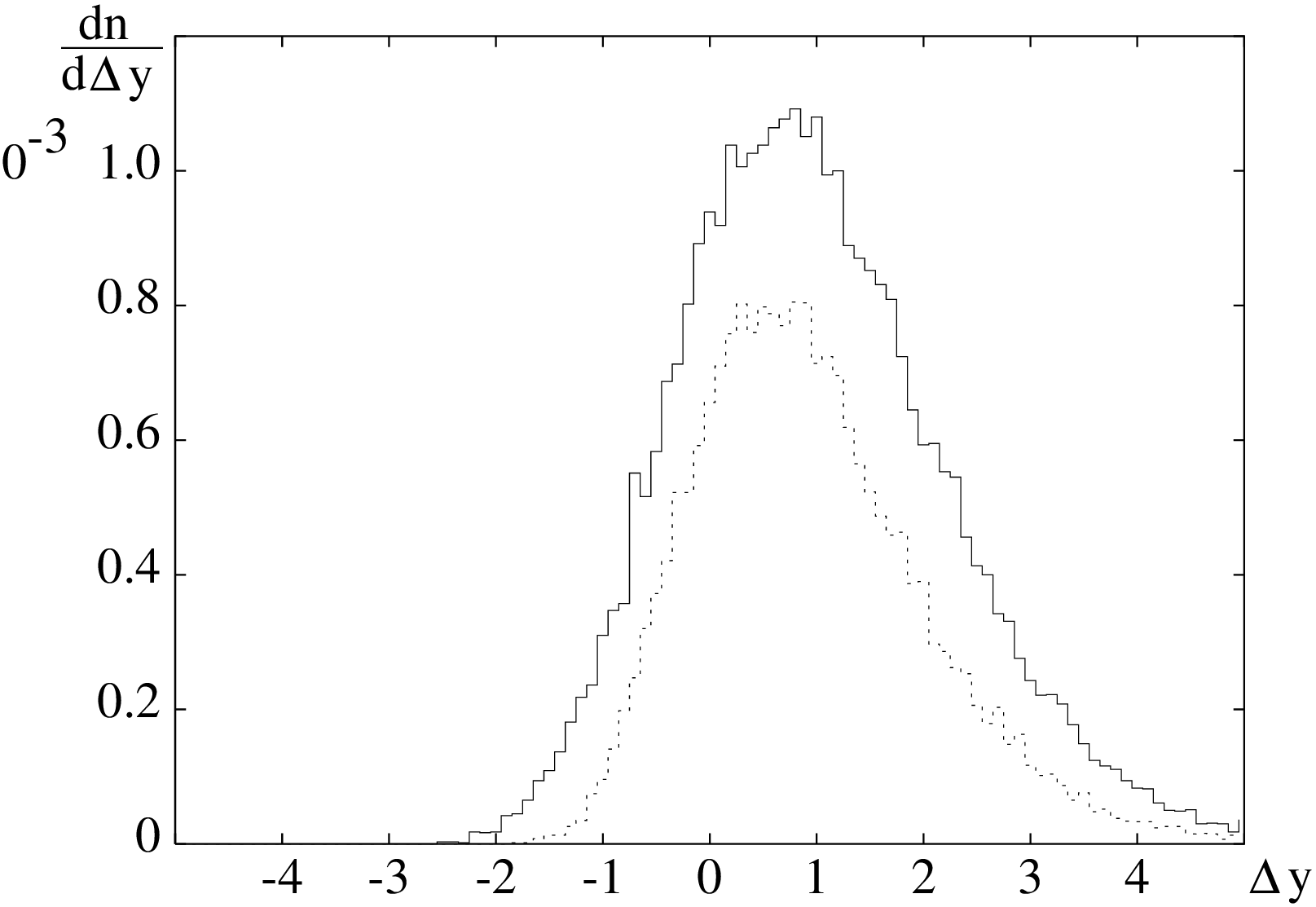,width=8cm}
          \psfig{figure=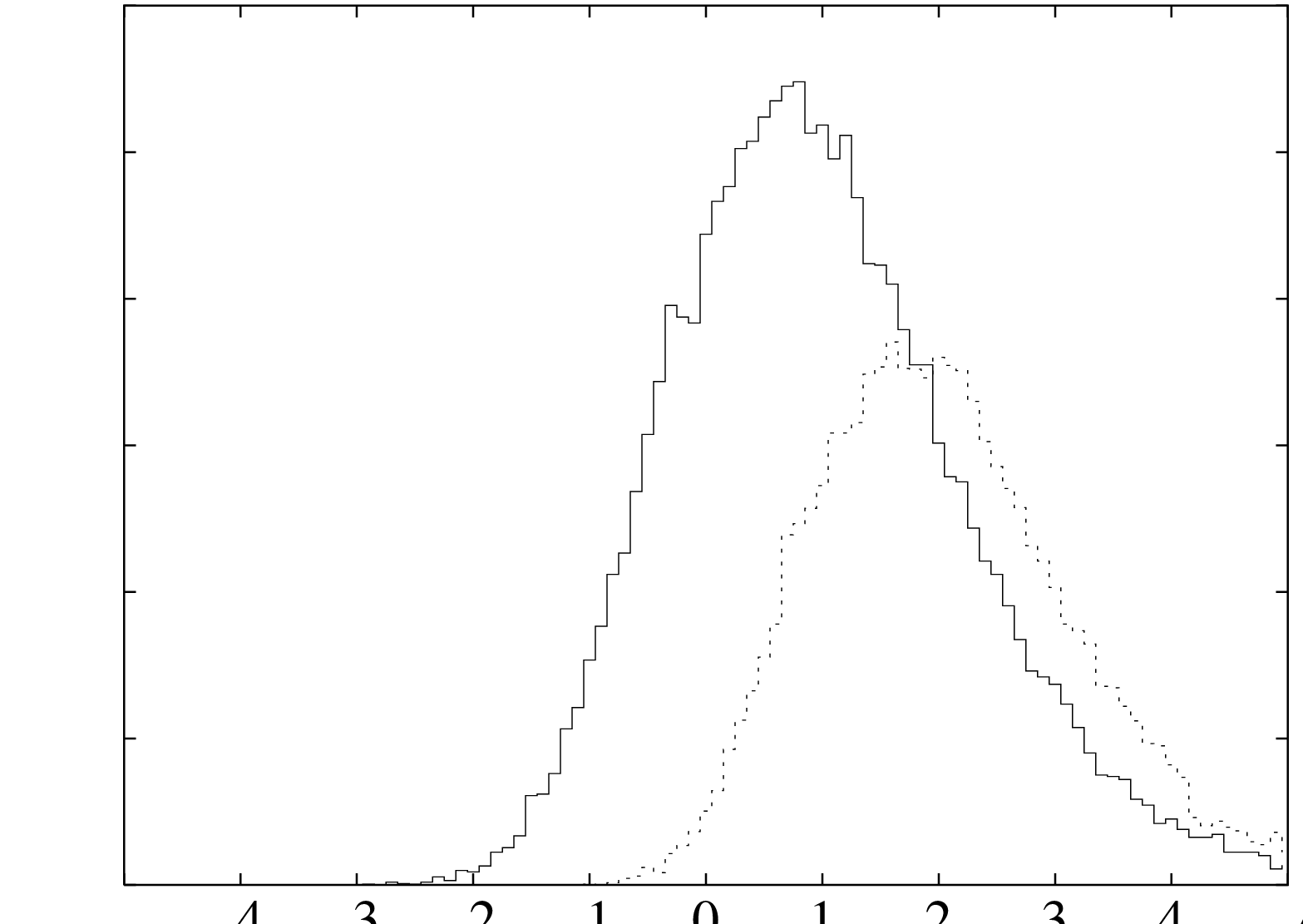,width=8cm}
	}
    \end{center}
  }}
  \caption{\em {\em a)} The full line shows the rapidity difference between rank {\em 2} and rank {\em 3} $\rho$'s, and the dotted line shows the rapidity difference between {\em B} and rank {\em 2} $\rho$'s, obtained by the modified splitting function, Eq~(\ref{e:modified}). {\em b)} Shows the same distributions when the original splitting function, Eq~(\ref{e:original}), is used. }
  \label{f:rapdiff}
\end{figure}

A relativistic string or a homogeneous (colour-)electric field is invariant under longitudinal boosts, which in the string model is reflected in a smooth distribution in rapidity for the produced hadrons. Although it is not at all a definite consequence, we feel that from the physical picture of hadrons formed from the energy stored in a string-like field, it is most natural that the distribution of light hadrons stretch up to the rapidity of the leading heavy B meson (or D meson in case of a c quark jet). Thus there should neither be a large rapidity gap nor should the rapidity distribution of the light hadrons continue beyond the rapidity of the heavy meson. This is actually the case for the splitting function in Eq~(\ref{e:modified}), while for the function in Eq~(\ref{e:original}) a large gap is obtained between the leading B-meson and the remainder of the jet. The average rapidity difference, $\overline{\Delta y}$, between the B meson and the first rank meson in the remainder is 0.88 which should be compared with the rapidity difference 0.91 between neighbouring light mesons in the middle of the string. The corresponding numbers for the harder distribution is 1.96 and 0.92, respectively. (As representative for a light meson we have used $\rho$ mesons to avoid kinematical effects from the exceptionally low pion mass, cf the discussion in~\cite{ba94}.) The difference between the two distributions is further illustrated in Fig~\ref{f:rapdiff}.
This figure shows the rapidity difference distributions between differently flavoured neighbouring mesons. The large (compared to the light mesons within the string) rapidity difference between the leading heavy B and the light $\rho$ is present for the hard distribution while the softer gives similar distributions irrespective to flavour and rank.

Such a smooth rapidity distribution is also expected in PQCD assuming LPHD or cluster fragmentation. This is a consequence of the fact that the initial b-quark radiates gluons with (pseudo-)rapidities up to the rapidity of the b-quark, but not beyond this value. The region of larger rapidities has been called the dead cone~\cite{yd91}.

\begin{figure}[t,b]
  \hbox{\vbox{
    \begin{center}
      \mbox{\psfig{figure=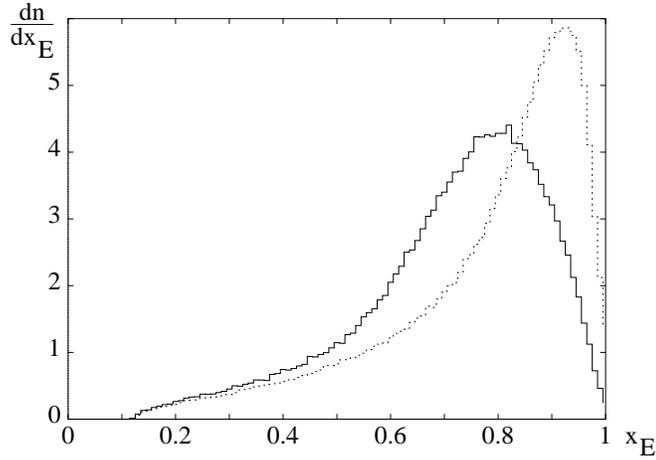,width=8cm}}
    \end{center}
  }}
  \caption{\em Combined $\mbn\mbp$ energy spectrum obtained at the Z pole, using the original (dotted line), and the modified (solid line) splitting functions, Eqs~(\ref{e:original}) and~(\ref{e:modified}), respectively. (Gluon radiation is included before fragmentation.)}
  \label{f:bspectrum}
\end{figure}
To be able to describe LEP $\epea \epe \ra \epq \epqa$ data correctly gluons must be included. In all simulations below gluons are radiated before string breakup. When we use the original symmetric Lund fragmentation formula (Eq~(\ref{e:original})) in the simulations we find that the average $\overline{x}_E$ for B-mesons decreases to 0.77. This is well outside the experimental value, $0.702\pm 0.002 \pm 0.008$ (average of the results reported from L3, ALEPH, OPAL and DELPHI experiments at LEP in 1994)~\cite{aa94}. Using the modified formula, Eq~(\ref{e:modified}), we find that $r_Q = 1.05 \pm 0.07$, gives an $\overline{x}_E$ within the experimental errors. We note that this is in very good agreement with the theoretically expected value 1. In Fig~\ref{f:bspectrum} we show the combined $\mbn$ and $\mbp$ energy spectrum using Eq~(\ref{e:original}) and Eq~(\ref{e:modified}), respectively. We see that the spectrum is considerably softened by the modified function.

The acceptable values of $r_Q$ shift down to 0.98$\pm 0.08$ when 30\% of the produced mesons have orbital angular momentum (tensor mesons) in the simulations. The B meson energy spectrum is softened by the (B$\pi$) resonance states, and in consequence the splitting function can be harder as compared with simulations without tensor mesons.

Using the softer b hadron energy distribution (Eq~(\ref{e:modified}) with $r_Q=1$) will increase the average number of charged particles produced in Jetset. The average difference between the number of charged particles produced in b events compared to uds events, $\overline{n}_{\mbox{\scriptsize b}} - \overline{n}_{\mbox{\scriptsize uds}}$, becomes~$3.06^{+0.29}_{-0.09}$ ($2.63^{+0.12}_{-0.26}$ for 30\% tensor mesons). This is in agreement with the experimental value, $3.14 \pm 0.44$~\cite{aa94}. Compared with an analytic MLLA calculation which gives $5.5 \pm 0.8$~\cite{bas92}, there is a large discrepancy. Despite this discrepancy we see that a detailed MC, with the same fundamental properties as MLLA, agrees with experimental data. The original splitting function in Eq~(\ref{e:original}), however, gives an average difference between the number of charged particles in b and uds events of 1.21, which is well below the experimental value.

Similar calculations have also been performed using other fragmentation models implemented in the Jetset MC: the Peterson \etal model for heavy quark fragmentation, inclusion of tensor meson production, and introduction of transverse momentum correlations and flavour correlations between neighbouring string breakups presented in~\cite{ba94}. In all simulations an optimization of the fragmentation parameters, $a$, $b$, and $\epsilon_b$, have been performed in order to try to retain several experimental observables ($\overline{n}_{ch}$, $\overline{n}_{\mbox{\scriptsize b}} - \overline{n}_{\mbox{\scriptsize uds}}$, $\overline{x}_E$ for B mesons, and average thrust) at LEP1 energies. In Table~\ref{t:tunings} we list results obtained when the different fragmentation models are used. The $a$ and $b$ parameters are chosen to give the best compromise for a given model, with the condition that the observed charged particle multiplicity at LEP1 is retained.
\begin{table}[t,b]
\begin{center}
\begin{tabular}{|c|ccccc|ccc|}
\hline
Splitting function	& Tensors & Corr. & a & b & $\epsilon_b$ & $\overline{n}_{\mbox{\scriptsize b}} - \overline{n}_{\mbox{\scriptsize uds}}$ & $x_E$ & $\left<\mbox{Thrust}\right>$ \\
\hline
Original Lund &0&no&0.30&0.49&-&1.2&0.77&0.93\\
Eq~(\ref{e:original}) &0.30&no&0.30&0.75&-&0.8&0.77&0.93\\
&0&yes&0.50&0.22&-&2.0&0.70&0.92\\
&0.30&yes&0.50&0.52&-&1.2&0.74&0.93\\
\hline
Modified Lund &0&no&0.30&0.58&-&3.1&0.71&0.93\\
Eq~(\ref{e:modified}) &0.30&no&0.30&0.90&-&3.0&0.72&0.93\\
&0&yes&0.90&0.44&-&3.3&0.66&0.93\\
&0.30&yes&0.90&0.90&-&3.0&0.69&0.93\\
\hline
Peterson \etal &0&no&0.30&0.53&0.0013&1.9&0.74&0.93\\
&0.30&no&0.30&0.85&0.0003&1.43&0.76&0.93\\
&0&yes&0.50&0.22&0.0082&2.4&0.68&0.92\\
&0.30&yes&0.50&0.57&0.0017&2.1&0.71&0.93\\
\hline
Experimental data &&&&&&3.14 \cite{aa94}&0.702 \cite{aa94}&0.9349 \cite{thrust}  \\
&&&&&&$\pm0.44$&$\pm0.002$&$\pm0.0006$\\
&&&&&&&$\pm0.008$&$\pm0.0024$\\
\hline
\end{tabular}
\end{center}
\caption{\em Some typical results obtained when the fragmentation models are tuned to retain the charged multiplicity at LEP1. The 'Tensor' column shows the fraction of tensor meson production in the simulations, and the 'Corr.' column shows if flavour correlations are included.}
\label{t:tunings}
\end{table}

We see that the Peterson \etal model, and the original splitting function, do not reproduce the experimental data. The modified splitting function performs well irrespective if tensor mesons are produced in the fragmentation or not, when no flavour correlations are included. However, in the presence of flavour correlations there seems to be a need of tensor meson production. We believe the fraction of tensor meson production used in the MC's are extreme values of the real fraction at LEP1, and note that the OPAL Collaboration expects the fraction to be at least 20\%~\cite{opal95}. Inclusion of flavour correlations in the simulations improve the performance of MC's when the Peterson \etal model or the original splitting function is used, but the discrepancy with experimental data is still quite large.

We conclude that the splitting function in Eq~(\ref{e:modified}) is both physically motivated and in better agreement with experimental data than the other fragmentation models implemented in Jetset, although these  cannot be fully ruled out on the basis of above considerations.

\section{B$\pi$ correlations as a flavour tag and CP violation} \vspace{-2ex}
In~\cite{mg92} it was suggested to use B$\pi$ correlations as a flavour tag in a study of CP violation in B decays. In this section we will study these correlations and estimate the efficiency in CP violation studies. We will also study how the relation between efficiency and purity varies with the mass of the B$\pi$ pair.

B$\pi$ correlations depend on the properties of b-jet fragmentation, and also on the production rate for B resonances decaying into a B$\pi$ pair. In the default version of the Jetset MC there are no correlations in flavour or transverse momentum between neighbouring $\epq \epqa$ string breakups. In~\cite{ba94} it is argued that due to the exceptionally small pion mass (the pions have also properties corresponding to a Goldstone boson) such correlations are expected to some degree. These flavour correlations could influence the B$\pi$ correlations, even if their effect on $\overline{x}_B$ and the inclusive spectra is negligible. In the following we will also study this possibility.

If a $\mbn$ $(\epba \epd)$ meson is produced in a $\epba$b event the associated $\epda$ is likely to end up in a $\mpip$ near in momentum space. This $\mpip$ could be part of the remaining jet, either as a directly produced pion or the decay product from e.g. a $\rho$ meson. The $\mbn \mpip$ pair can also be decay products from a parent B resonance. From iso-spin invariance the $\mbn \mpip$ correlations are related to B$^\pm \pi$ correlations, which are experimentally studied by the OPAL collaboration~\cite{opal95}.

Let us first study the case where the $\epda$ partner of a $\mbn$ is a leading parton in the remaining jet. The tagging possibility relies on the fact that the excess positive charge from the $\epda$ is found in the leading end of the jet, usually in the first rank hadron (of the remaining jet). If it is found in the first rank hadron the characteristic invariant B$\pi$ mass is determined by the relation
\eqbe
M^{char}_{\mbox{\scriptsize B}\pi} \approx {M_{\mbox{\scriptsize B}}} + m_{\perp} \cosh \Delta'
\label{e:charmass}
\eqen
where $m_\perp$ is the transverse mass of the pion and $\Delta'$ is the rapidity difference between the B and the pion or the parent resonance if the pion comes from e.g. a $\rho$ meson. Since the energy spectrum for the B mesons is strongly peaked at high energies, with an average around $\overline{x}_E\approx0.7$, the B$\pi$ mass increase rapidly for higher rank pions with lower energy in the cms. Thus to enhance the signal it is favourable to study only pairs with mass below 5.8GeV. We also note that low energy B mesons can be produced in perturbative gluon splittings, $\ggn\ra\epb\epba$, but this is estimated to be a very small contribution at LEP1 energies~\cite{hp96}.

In ref~\cite{opal95} it is assumed that the next $\epq\epqa$ pair in the string breakup has the same probability to be a $\epu\epua$ as a $\epd\epda$ pair. This is not necessarily the case, even if the fragmentation is iso-spin invariant. Consider a toy model in which only pions are produced. Then iso-spin invariance implies equal numbers of $\pi^+$, $\pi^0$, and $\pi^-$, which means that a $\epd\epda$ pair must be followed by a $\epu\epua$ with probability 2/3 and a $\epd\epda$ only with probability 1/3. (Equal probabilities would give twice as many $\pi^0$ as $\pi^+$ or $\pi^-$.) This is the type of flavour correlations discussed in ref~\cite{ba94}.

From our estimates the effect of a realistic degree of flavour correlations is small. It could conceivably affect an estimate of B$\pi$ resonance by distorting the background. The flavour correlations imply that the excess positive charge is over compensated in the first rank meson, followed by a negative contribution in the second rank, a still smaller positive contribution in the third rank meson, etc. We want in the future to estimate the magnitude of such a possible distortion, to see if it could have a noticeable affect on the experimental results. We note that the OPAL Collaboration estimates that (B$\pi$) resonances contributes with at least 20\% of the total production of B mesons in $Z^0 \ra \epb \epba$ fragmentation~\cite{opal95}. 

In~\cite{mg92} a mass estimate for the (B$\pi$) resonance is obtained from extrapolation of charmed meson data to B mesons. The estimate puts an upper limit on the resonance mass at 5.8GeV. This estimate is supported by OPAL data where an excess, as compared with simulations where no resonance meson production was included, of $\mbox{B}^+ \pi$ pairs is found at 5.7GeV. The (B$\pi$) resonance will eventually decay to a B$\pi$ pair close in phase space, and contribute to this excess.

Pions produced in the center of the string have a fairly large momentum relative to the $\mbn$ meson. If we want to retain $\mbn \pi$ pairs close in phase space, the relative momentum between central pions and the B puts kinematical constraints on the B energy, forcing the B energy to be small. We have seen above that the average $x_E$ for B mesons is 0.70, so central pions will have small chances of making a small mass pair with the B. Low $x_E$ mesons can 

The method to trace CP violating B mesons proposed in ref~\cite{mg92} depends on the correlation of $\mbn$ ($\mbna$) mesons with pions nearby in the phase space, {\em and} the detection of a CP eigenstate f as a $\mbn$ or $\mbna$ decay product in conjunction with a similar pion.

The suggested method leads to a dilution of {\em any} real CP violating asymmetry. Starting with the time integrated asymmetry~\cite{id86}
\eqbe
A(f) \equiv \frac{\Gamma(\mbn_{t=0} \ra \ferm) - \Gamma(\mbna_{t=0} \ra \ferm)}
	         {\Gamma(\mbn_{t=0} \ra \ferm) + \Gamma(\mbna_{t=0} \ra \ferm)},
\eqen
where f denotes a CP eigenstate and $\Gamma(\mbn_{t=0}\ra\ferm)$ is the decay width when the state f is produced as a decay product of a state starting out as a $\mbn$ at t=0.

Expressing the asymmetry in final states we get
\eqbe
A(f) = \frac{1}{1+x^2} 
       \frac{[N(\overline{\mbox{T}} \mpim) + N(\mbox{T} \mpim)]
			N_{\mbox{\scriptsize \ferm}^+} - 
	     [N(\mbox{T} \mpip) + N(\overline{\mbox{T}} \mpip)]
			N_{\mbox{\scriptsize \ferm}^-}}
	    {[N(\overline{\mbox{T}} \mpim) - N(\mbox{T} \mpim)]
			N_{\mbox{\scriptsize \ferm}^+} + 
	     [N(\mbox{T} \mpip) -N(\overline{\mbox{T}} \mpip)]
			N_{\mbox{\scriptsize \ferm}^-}}.
\eqen
T is a flavoured state we know to come from a $\mbn$, and is used to determine the flavour of the B meson. $N(\mbox{T} \pi)$ is the relative number (probability) of states T in conjunction with a pion. $N_{\mbox{\scriptsize \ferm}^\pm}$ denotes the number of final states $\ferm \pi^\pm$. $x$ is the mass mixing parameter $x \equiv (\Delta m/\Gamma)$. (Notation taken from~\cite{mg92}.)

Considering a charge symmetric production process ($\hp \hpa$ or \epem) where $\mbn$ and $\mbna$ production should be equal, ie. $N(\mbox{T} \mpim) = N(\overline{\mbox{T}} \mpip)$ and $N(\mbox{T} \mpip) = N(\overline{\mbox{T}} \mpim)$, gives
\eqbe
A_{obs} \equiv
	\frac{N_{\mbox{\scriptsize \ferm}^+}-N_{\mbox{\scriptsize \ferm}^-}}
	     {N_{\mbox{\scriptsize \ferm}^+}+N_{\mbox{\scriptsize \ferm}^-}} =
	\frac{N(\mbox{T} \mpip) - N(\mbox{T} \mpim)}
	     {N(\mbox{T} \mpip) + N(\mbox{T} \mpim)}
	(1 + x^2) A(f),
\label{e:aobs}
\eqen
and we see that the tagging process suppresses the experimentally observed asymmetry. We are going to estimate the efficiency of the suggested tagging method by calculating the dilution factor,
\eqbe
\xi \equiv \frac{N(\mbox{T} \mpip) - N(\mbox{T} \mpim)}
		{N(\mbox{T} \mpip) + N(\mbox{T} \mpim)},
\label{e:dilution}
\eqen
of Eq~(\ref{e:aobs}) for the reaction \epem $\ra \epb \epba \ra$ X as a function of the B$\pi$ pair mass. From the B production considerations above we expect an excess of low mass $\mbn \mpip$ pairs giving a non vanishing dilution factor.

\begin{figure}[t,b]
  \hbox{\vbox{
    \begin{center}
       \mbox{
	\psfig{figure=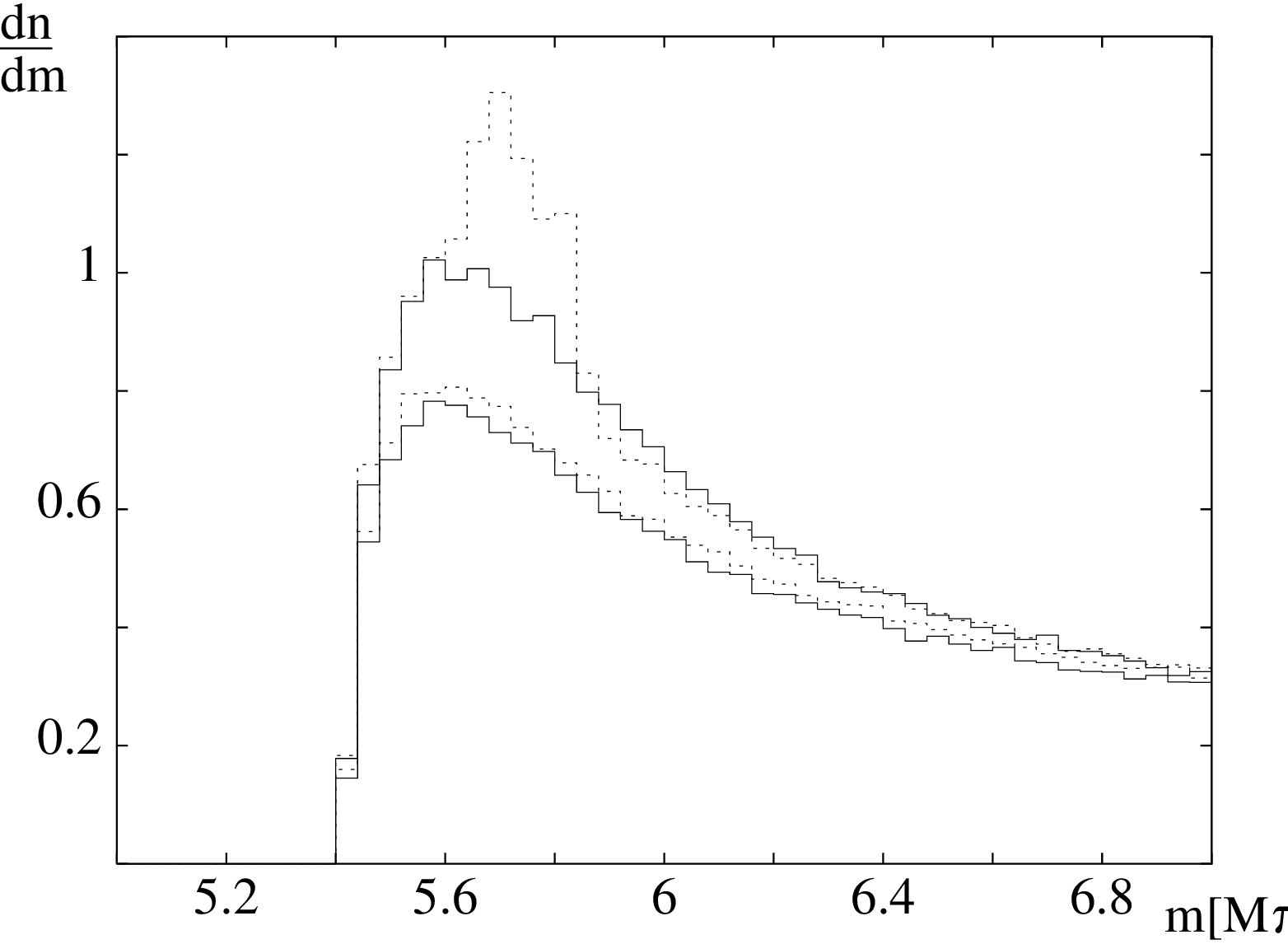,width=8cm}
	\psfig{figure=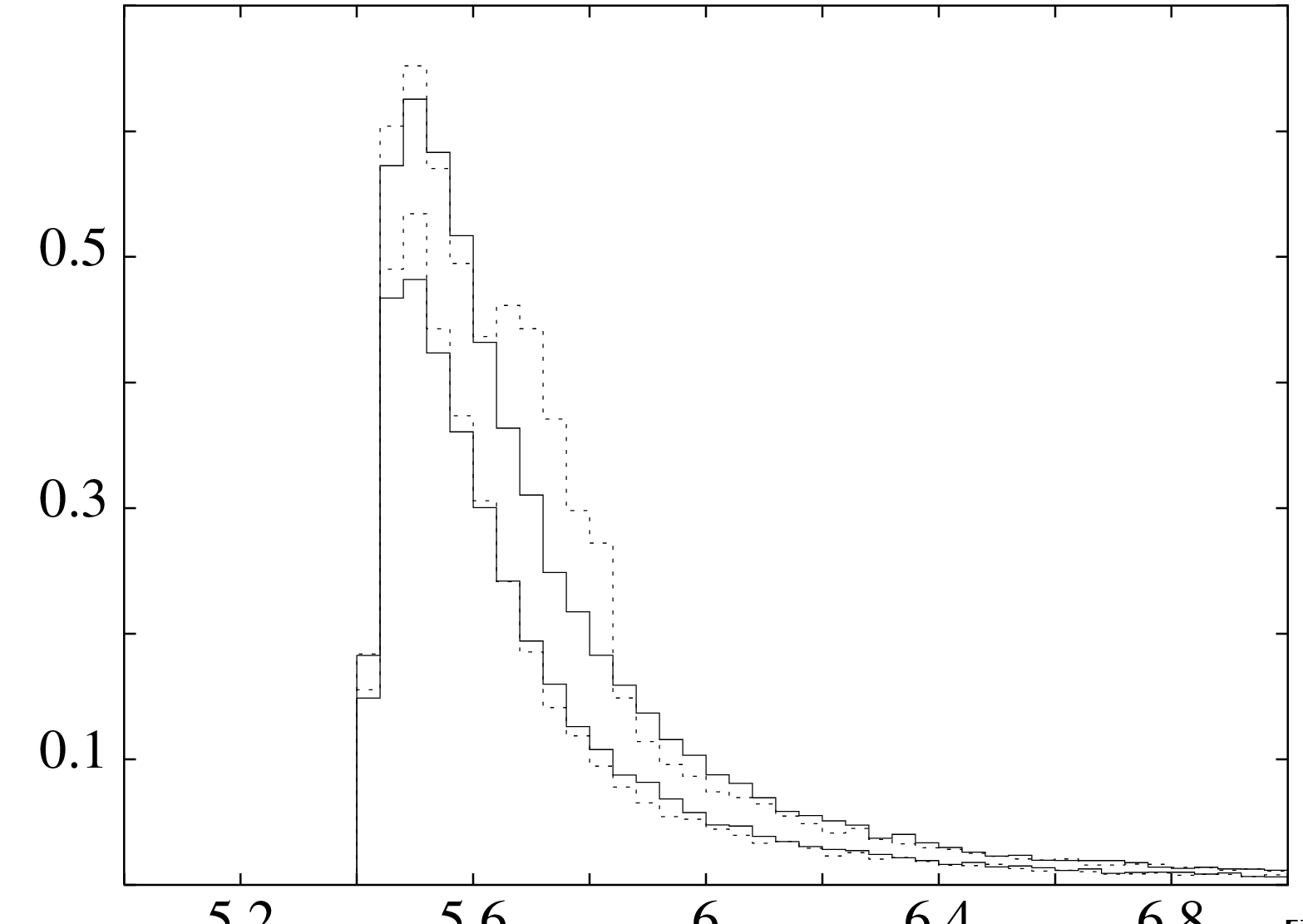,width=8cm}
	}
    \end{center}
  }}
  \caption{\em Invariant mass multiplicity distributions. \protect\newline
{\bf a}) Events where all possible $\mbn \pi$ pairs are considered. The full lines are data from events without any tensor mesons. The upper full line is for $\mbn \mpip$ pairs, and the lower full line is for $\mbn \mpim$ pairs. The dashed lines are for events with a tensor meson fraction of 30\%. Again the upper (dashed) line is for $\mbn \mpip$ and the lower for $\mbn \mpim$. \protect\newline
{\bf b}) Events where only the fastest pion is paired with the B meson. The different lines describe the same sort of data as in a).}
  \label{f:invmass}
\end{figure}

In the calculations we use Jetset with the softer fragmentation distribution, Eq~(\ref{e:modified}), since this is in best agreement with experimental data. As the tagged state we use the $\mbn$ meson and accept all possible $\mbn \pi$ pairs. In Fig~\ref{f:invmass}a we show the invariant mass multiplicity distribution for $\mbn \pi$ pairs from $10^6$ mixed flavour events at LEP1 energies. Of these approximately 218000 events will be $\epb \epba$ events giving b hadrons. In the figure we clearly see a difference between the $\mbn \mpip$ and $\mbn \mpim$ distributions. Continuing to large $m($B$\pi)$ the difference decreases since the B$\pi$-charge correlations vanishes.

In the calculation of the average dilution factor we only include B$\pi$ pairs with masses below 5.8GeV. This restriction eliminates the chance of using central pions in the pair and emphasize the contributions from (B$\pi$) resonances which are more probable to give a right charged pion. Accepting all possible pairs with $m(\mbn \pi) \leq 5.8$GeV the average dilution factor is $\xi=$0.13--0.17. The variation in $\xi$ is mainly due to different production fractions of tensor mesons allowed in the simulations, and we note that a larger fraction of resonance meson production will enhance the efficiency of the tagging method.

The efficiency of the proposed tagging method will increase if only the most correlated pion in an event is taken into account. This pion should be the fastest one in the event. In Fig~\ref{f:rapdiff} we have plotted the rapidity difference between rank 2 and rank 3 mesons in the Lund string. Indeed we see that in general the fastest meson is the rank 2 particle, ie. the neighbour to the b hadron. (In Fig~\ref{f:rapdiff} we show distributions for $\rho$ mesons, but similar results are obtained for any directly produced particles.) Considering only this pion in every event gives a more peaked invariant mass distribution as can be seen in Fig~\ref{f:invmass}b. The average dilution factor, for pairs with $m($B$\pi)\leq5.8$GeV, increases to 0.17--0.20. Again the dilution factor depends on the fraction of tensor mesons produced in the simulations.

\begin{figure}[t,b]
  \hbox{\vbox{
    \begin{center}
       \mbox{
	\psfig{figure=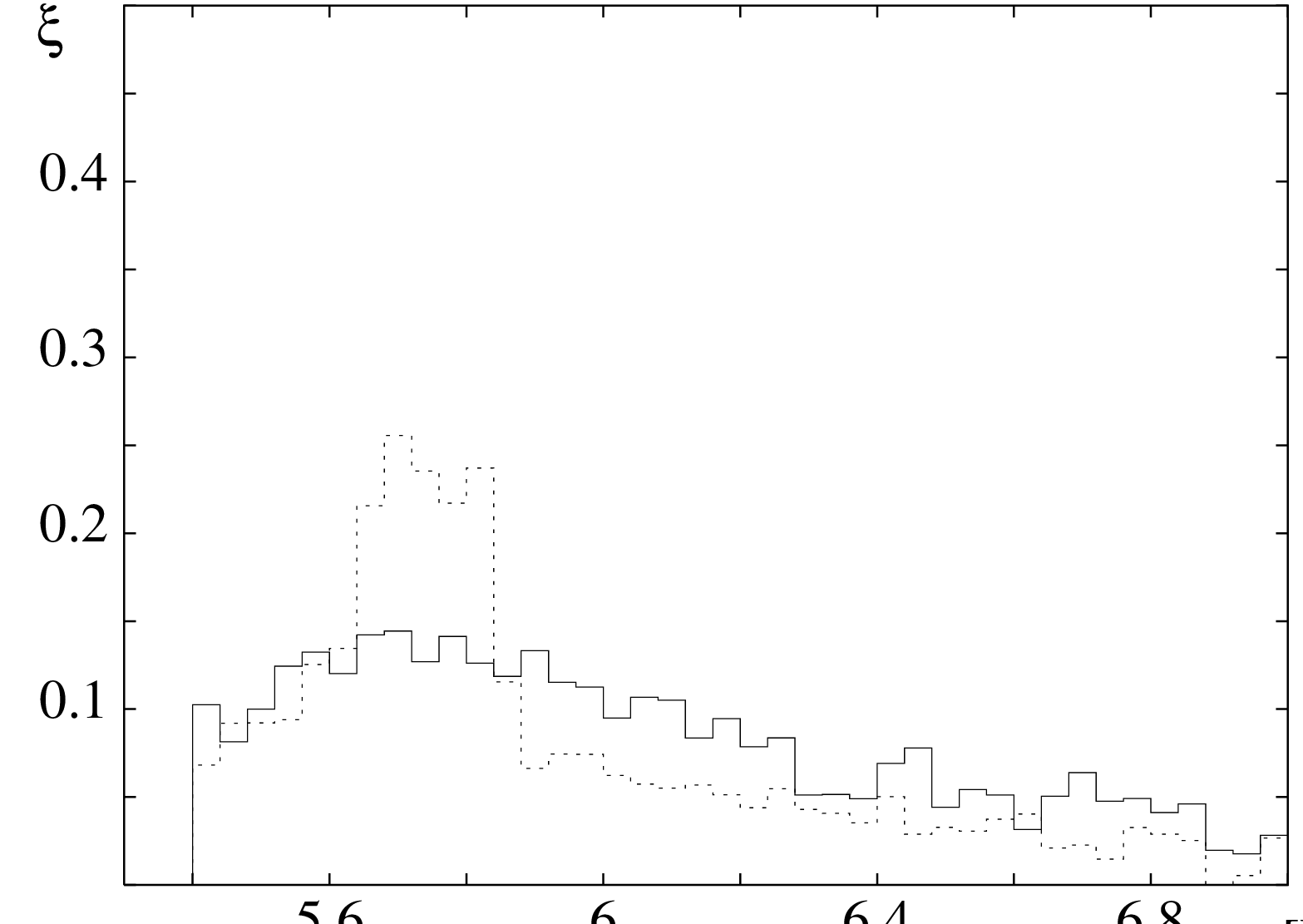,width=8cm}
	\psfig{figure=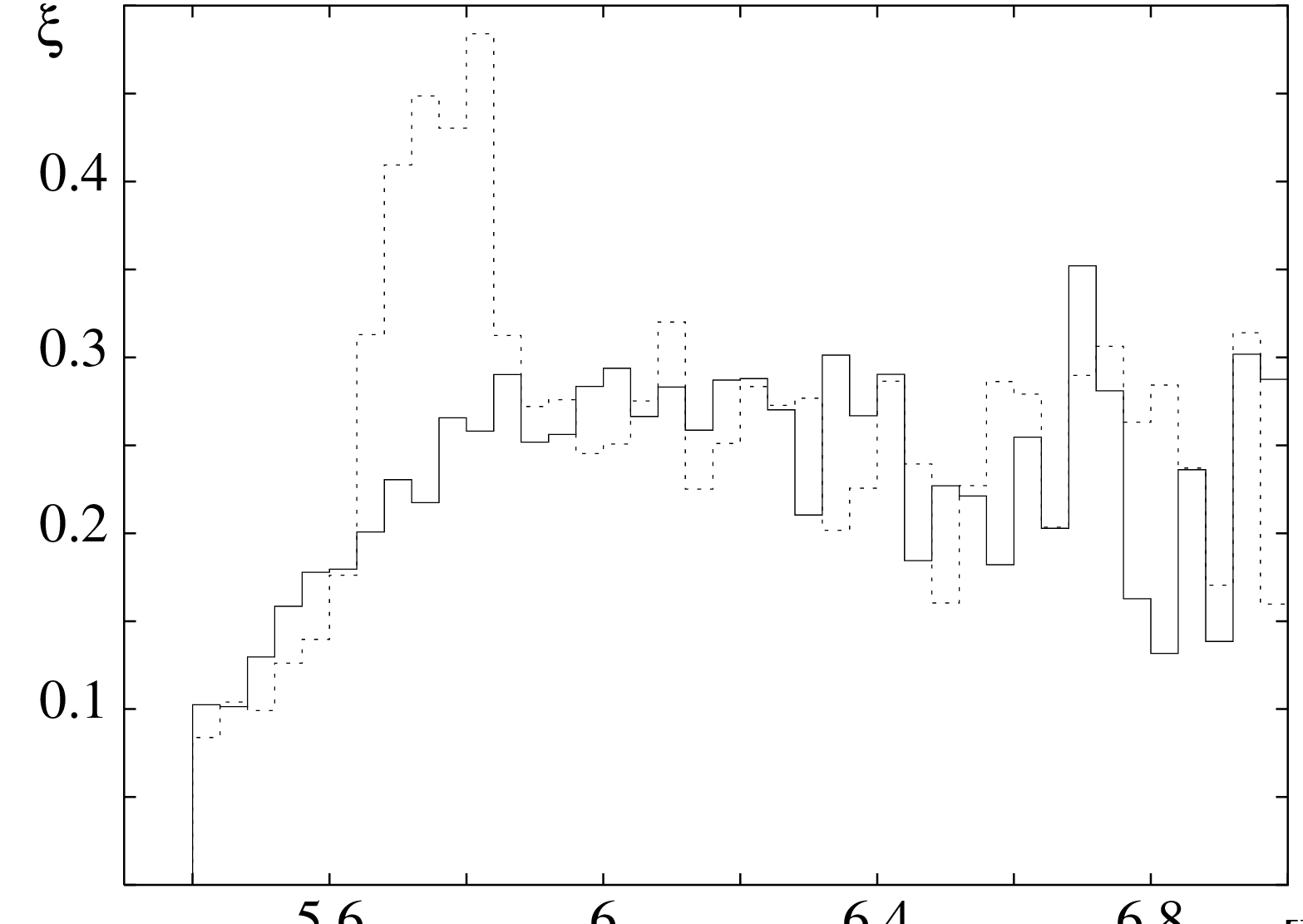,width=8cm}
	}
    \end{center}
  }}
  \caption{\em Dilution factor as a function of invariant mass. \protect\newline
{\bf a}) Events where all possible $\mbn \pi$ pairs are considered. The full line is for events without any tensor mesons, and the dashed line is for events with a tensor meson fraction of 30\%. \protect\newline
{\bf b}) Events where only the fastest pion is paired with the B meson. The two lines describe the same sort of data as in a).}
  \label{f:correlation}
\end{figure}

The distributions in Fig~\ref{f:invmass} can be used to calculate the dependence of the dilution factor on the B$\pi$ invariant mass. The result is plotted in Fig~\ref{f:correlation}. We see that at $m($B$\pi)=5.5$GeV the dilution factor is about 0.10 and grows to a maximum at $m($B$\pi)=5.8$GeV. The growth is much larger if only the fastest pion is used as can be seen in Fig~\ref{f:correlation}b. The large difference in the maximum of $\xi=$ around 5.8GeV depend on whether tensor mesons are produced or not, since the decay of B$^{**}$ mesons contributes to the $\mbn \mpip$ and therefore enhance the dilution factor. In Fig~\ref{f:correlation}b $\xi$ remains around 0.25 above m(B$\pi)=6$GeV, but very low rates are expected above this mass, cf. Fig~\ref{f:invmass}b.

When the tensor meson production rate is assumed to be within 20--30\%, we calculate the dilution factor in the invariant mass range $5.6<m\left(\mbn\pi\right)<5.9$GeV to be 0.31--0.35. The OPAL Collaboration finds a similar value, 0.364$\pm0.024$, in their $\mbp \pi$ studies~\cite{opal95}, where it is expected that at least 20\% of the B mesons are produced as decay products of heavier resonances at LEP1

We have also performed the same calculations with an MC where we included flavour correlations between neighbouring $\epq\epqa$ pairs. The results from these runs are very similar to those found above without such flavour correlations.

\section{Conclusions}
A modification of the original Lund string fragmentation function (Eq~(\ref{e:original})), where the reduction in the ``colour coherence area'' due to the large mass of heavy quarks is taken into account, will give a softer energy distribution for heavy mesons. Experimental data are reproduced by tuning the parameter $r_Q$ to $1.05 \pm 0.07$ (cf. Eq~(\ref{e:modified})), which is consistent with the theoretically derived $r_Q \equiv 1$. The Lund string model, together with the modified splitting function, also gives rapidity distributions in agreement with PQCD assuming LPHD or cluster fragmentation. The Jetset MC is found to perform better when the modified splitting function is used, as compared to the the original splitting function or the Peterson \etal fragmentation model.

The possibility to experimentally observe CP violating B mesons using pion charge to tag the flavour of the leading $\mbn$ is always diluted. This dilution depends on the fraction of resonance B meson production. In the OPAL measurement~\cite{opal95} it is found that at least 20\% of the B mesons are produced as decay products of heavier resonances at LEP1. We estimate the dilution factor, defined in Eq~(\ref{e:dilution}), in the invariant mass range, $5.6<m\left(\mbn\pi\right)<5.9$GeV to be 0.31--0.35. The uncertainty is due to the fraction of resonance production, here we have used 20--30\%. The OPAL Collaboration finds a similar value, 0.364$\pm0.024$, in their $\mbp \pi$ studies~\cite{opal95}.

\noindent {\bf Acknowledgments} \\
I would like to thank prof G\"osta Gustafson and dr Torbj\"orn Sj\"ostrand for valuable discussions. 


\vspace{-2ex}

\begin{thebibliography}{99} \vspace{-1ex}
  \bibitem{mg92}
	\bibl{M. Gronau, A. Nippe, J.L. Rosner}{Phys. Rev.}{D47}{1993}{1988}
  \bibitem{ba83}
	\bibl{B. Andersson, G. Gustafson, G. Ingelman, T. Sj\"ostrand}
		{Phys. Rep.}{97}{1983}{31}
  \bibitem{ts94}
	\bibl{T. Sj\"ostrand}{Comp. Phys. Comm.}{82}{1994}{74}
  \bibitem{bw84}
	\bibl{B.R. Webber}{Nucl. Phys.}{B238}{1984}{492} \\
	\bibl{G. Marchesini, B.R. Webber}{Nucl. Phys.}{B238}{1984}{1}
  \bibitem{gm92}
	\bibl{G. Marchesini {\em et al.}}{Comp. Phys. Comm.}{67}{1992}{465}
  \bibitem{cp83}
	\bibl{C. Peterson, D.Schlatter, I. Schmitt, P. Zerwas}
		{Phys. Rev.}{D27}{1983}{105}
  \bibitem{mb81}
	\bibl{M.B. Bowler}{Z. Phys.}{C11}{1981}{169} \\
	\bibl{D.A. Morris}{Nucl. Phys.}{B313}{1989}{634}
  \bibitem{ba83:2}
	\bibl{B. Andersson, G. Gustafson, and B. S\"oderberg}
		{Z. Phys.}{C20}{1983}{317}
  \bibitem{ba94}
	\bibl{B. Andersson, G. Gustafson, J. Samuelsson}
		{Z. Phys.}{C64}{1994}{653}
  \bibitem{yd91}
	\bibl{Yu.L. Dokshitzer, V.A. Khoze, S.I. Troyan}
		{J. Phys.}{G17}{1991}{1602}
  \bibitem{aa94}
	A. De Angelis, {\em CERN-PPE/94-174}
  \bibitem{bas92}
	\bibl{B.A. Schumm, Yu.L. Dokshitzer, V.A. Khoze, D.S Koetke}
		{Phys. Rev. Lett.}{69}{1992}{3025}
  \bibitem{thrust}
	\bibl{The L3 Collab}{Z. Phys}{C55}{1992}{39}
	\bibl{The ALEPH Collab}{Z. Phys}{C55}{1992}{209}
	\bibl{The DELPHI Collab}{Z. Phys}{C59}{1993}{21}
	\bibl{The OPAL Collab}{Z. Phys}{C59}{1993}{1}
  \bibitem{opal95}
	\bibl{The OPAL Collab.}{Z. Phys.}{C66}{1995}{19}
  \bibitem{hp96}
	Helenka Przysiezniak-Frey, Talk given at Moriond: QCD and High Energy Hadronic Interactions 1996
  \bibitem{id86}
	\bibl{I. Dunetz, J.L. Rosner}{Phys. Rev.}{D34}{1986}{34} \\
	\bibl{I. Dunetz}{Ann. Phys. (N.Y.)}{184}{1988}{350}
\end{thebibliography}
\end{document}